\documentclass{article_saj}
\usepackage{graphicx,saj,multicol,subeqnarray}
\usepackage{natbib}
\usepackage{float}
\usepackage{xcolor}
\usepackage{widetext}
\usepackage{url}
\usepackage{bm}
\usepackage{tikz} 
\usepackage{pifont} 
\usepackage{amsfonts}
\usepackage{amssymb}
\usepackage{amsmath,upgreek}
\usepackage{titlesec}
\titlelabel{\thetitle.\quad}
\definecolor{xlinkcolor}{cmyk}{1,0.6,0,0}
\usepackage[bookmarks=false,         
     pdfnewwindow=true,      
     colorlinks=true,    
     linkcolor=xlinkcolor,     
     citecolor=xlinkcolor,     
     filecolor=xlinkcolor,  
     urlcolor=xlinkcolor,      
final=true
]{hyperref}
\usepackage{changepage}
\usepackage{diagbox}
\usepackage{stackengine}

\def\udc{52}

\begin{document}
\parindent=.5cm
\baselineskip=3.8truemm
\columnsep=.5truecm
\newenvironment{lefteqnarray}{\arraycolsep=0pt\begin{eqnarray}}
{\end{eqnarray}\protect\aftergroup\ignorespaces}
\newenvironment{lefteqnarray*}{\arraycolsep=0pt\begin{eqnarray*}}
{\end{eqnarray*}\protect\aftergroup\ignorespaces}
\newenvironment{leftsubeqnarray}{\arraycolsep=0pt\begin{subeqnarray}}
{\end{subeqnarray}\protect\aftergroup\ignorespaces}
%


\markboth{\eightrm Red Nova Progenitor} 
{\eightrm S. WADHWA {\lowercase{\eightit{et al.}}}}

\begin{strip}

{\ }

\vskip-1cm

\publ


{\ }


\title{KIC 9272276: An Extreme Low Mass Ratio Red Nova Progenitor }


\authors{S. Wadhwa$^{1}$, M. Grozdanović$^{2}$, B. Arbutina$^{3}$, N.F.H Tothill$^{1}$, M.D Filipovi{\' c}$^{1}$ and  A.Y De Horta$^{1}$}

\vskip3mm


\address{$^1$School of Science, Western Sydney University,\break Locked Bag 1797, Penrith, NSW 2751, Australia.}


\Email{19899347@student.westernsydney.edu.au}

\address{$^2$Astronomical Observatory, Volgina 7, 11060 Belgrade 38, Serbia}

\address{$^3$Department of Astronomy, Faculty of Mathematics, University of Belgrade, Studentski trg 16, 11000 Belgrade, Serbia.}






\summary{Photometric analysis of KIC 9272276 (K9272), a low-mass contact binary system, is presented. We find it to be an example of an extreme low mass ratio system that satisfies the current theoretical criteria for a potential red-nova progenitor. V band photometry of the system was modelled using the Wilson-Devenney code and we find the system to be in marginal contact with a mass ratio of 0.081. The estimated mass of the primary component is one solar mass and the theoretical instability mass ratio range, accounting for both metallicity and age, is well above the modelled mass ratio at 0.09 - 0.101. We find no evidence of a potential third light contamination and discuss other findings confirming our analysis.}



\end{strip}

\tenrm


\section{Introduction}

\indent

\footnotetext[0]{\copyright \ 2023 The Author(s). Published by
Astronomical Observatory of Belgrade and Faculty of Mathematics,
University of Belgrade. This open access article is distributed under
CC BY-NC-ND 4.0 International licence.}

W Ursa Majoris binary systems usually consist of a low mass (spectral class F to K) main-sequence primary \citep{2013MNRAS.430.2029Y} with an even lower mass secondary. Due to their proximity both components are tidally distorted such that the larger primary expands beyond its Roche lobe while still on the main sequence and there is significant transfer of mass from the primary to the secondary. The smaller secondary is not able to accrete the mass inflow, and it also expands beyond its Roche lobe, leading to bidirectional mass exchange, and the system becomes enclosed in a co-rotating common envelope. The rate of mass transfer depends on the initial masses of the components and how quickly the primary expands beyond its Roche lobe. Based on the model presented by \citet{2008MNRAS.390.1577G} and an initial pre-contact mass ratio of 0.5, a solar mass primary would lose nearly half its mass over an 8-9Gyr period to the secondary prior to the start of the contact phase. There is considerable transfer of energy from the primary to the secondary by processes not fully elucidated such that the temperature of the common envelope is close to that of the primary component and the secondary component is considerably more luminous than its main sequence counterpart of the same mass.

The current consensus is that angular momentum loss (AML) through magnetic breaking plays a significant role in contact binary evolution \citep{1981ApJ...245..650M, 1981Ap&SS..78..401V, 2004MNRAS.355.1383L}. Contact binary orbit can remain synchronised even with progressive AML up until some critical separation where tidal instability (also known as Darwin instability) sets in \citep{1879Obs.....3...79D}. At this juncture, the orbital and rotational (spin) momentum approaches a 1:3 ratio, and the system can no longer maintain a synchronous orbit \citep{1980A&A....92..167H}. With continued AML, the orbit will begin to shrink rapidly, leading to the likely ejection of the common envelope and merger of the components. The Component merger is postulated to lead to a bright transient known as a red nova. However, to date there is only one confirmed red nova event secondary to contact binary merger that of V1309 Sco \citep{2011A&A...528A.114T}. 

Modelling has always suggested that merger would take place when the component mass ratio ($M_2/M_1$) ($q$) is extremely low \citep{1995ApJ...444L..41R, 2009MNRAS.394..501A}. Earlier considerations adopted a potential "universal" minimum mass ratio below which contact binaries would merge \citep{1995ApJ...438..887R, 2007MNRAS.377.1635A, 2009MNRAS.394..501A}. Recent theoretical updates have shown that there is unlikely to be a single minimum mass ratio at which orbital instability would begin. It is now suggested that each system has its unique mass ratio at which it is likely to become unstable \citep{2021MNRAS.501..229W, 2024SerAJ.208....1A}. In the case of low mass contact binaries, with primaries between $0.6\rm M_{\odot}$ and $1.4\rm M_{\odot}$, the instability mass ratio can range from 0.04 to 0.22. In addition, there is evidence that the internal composition (metallicity) and the age of the contact binary, both of which have a sufficient influence on the internal gas dynamics of the star to influence its gyration radius, also have a significant influence on orbital stability \citep{2024MNRAS.535.2494W}. The net effect of the recent updates suggests that most contact binaries in the Solar vicinity are likely to be stable. \citet{2024MNRAS.535.2494W} were unable to confirm any unstable system from hundreds of published light curve solutions.

As part of our ongoing search for potential merger candidates, we have used the rapid identification procedure from \citet{2022JApA...43...94W} to identify K9272 (= ASAS-SN-V J191059.37+454316.6, T-Lyr-07958) as a potential red nova progenitor from the All Sky Automated Survey - SuperNova (ASAS-SN) \citep{2020MNRAS.491...13J}. The identification procedure links the estimated mass of the primary component (derived through colour and survey spectra) and the amplitude of the observed light curve with the theoretical instability mass ratio. If the amplitude is significantly higher than the expected amplitude for the instability mass ratio, than the system can be considered stable without further analysis. K9272 was recognised as variable by the Automated Survey of the Kepler mission fields in 2008 \citep{2008arXiv0808.2558P}. It has been observed as part of both the Kepler and TESS missions as well as ASAS-SN. Photometry from all these surveys demonstrate total eclipses, indicating the system to be suitable for photometric analysis, and low amplitude, suggesting possible mass ratio in the instability range. We report the modelling of V band photometry and orbital stability analysis confirming that the system is potentially unstable based on the updated instability criteria.

\section {Observations and Mass of the Primary Component}

K9272 was observed over 8 nights between July 2024 and April 2025 with the 0.4m telescopes from the Las Cumbres Observatory (LCO) network (SBIG STL-6303 CCD camera) and 0.6m telescope (FLI ProLine PL23042 CCD camera) at the Astronomical Station Vidojevica in Serbia. Full phase light curve was obtained in the $V$ band (Bessell) and to document the colours of the primary we also acquired high cadence images during eclipses in the $B$, $V$ and $R$ bands (Johnson-Cousins). All images were calibrated with appropriate dark, bias and flat field frames. We used the AstroImageJ \citep{2017AJ....153...77C} package to perform differential photometry with UCAC4 679-066810 (B: 14.29, V: 12.97 and R: 12.71 magnitudes from \citet{2015AAS...22533616H}) as the comparison star and UCAC4 679-066805 as the check star. The software package calculates the difference in apparent magnitude between the comparison star and the target star and based on the reported apparent magnitude of the comparison star provides the apparent magnitude of the target star.

We used our data and the available data from ASAS-SN, Kepler, TESS and our observations to refine the ephemeris as follows:
\begin{center}
    
    $BJD_{min} = 2460781.0830844(\pm221) + 0.28061529(\pm200)E$.\\                   

\end{center}

Our light curve demonstrates a $V$ band amplitude of 0.204 mag which is similar to the ASAS-SN amplitude of 0.21 mag, Kepler amplitude of 0.193 magnitude. The TESS amplitude is slightly higher at 0.226 mag. The TESS flux, obtained through from The Mikulski Archive for Space Telescopes (MAST) data portal, was converted to magnitude using the relationship $-2.5log10(flux) + 20.44$ from the TESS instrument handbook. Our observations indicate the following colours for the primary component $B-V = 0.97$ and $V-R = 0.13$. The light curve shows both maxima of equal brightness and similarly, both minima are of equal depth. The apparent $V$ band magnitude range is 13.154 - 13.358. The infinite-length extinction is estimated as 0.0591 \citep{2011ApJ...737..103S} and corrected for distance \citet{2023RAA....23k5001W} as 0.044, yielding an extinction magnitude of 0.138 mag. As the secondary eclipse is complete, the observed light during this time represents the light from the primary component, and we estimate its extinction corrected apparent magnitude as 13.22, which when combined with \textit{GAIA} DR 3 \citep{2016A&A...595A...1G, 2023A&A...674A...1G} estimated distance of 640.09 pc yields an absolute magnitude ($M_{V1}$) of 4.46 for the primary component and 4.26 ($M_{VT}$) for the system at maximum brightness indicating an absolute magnitude of 6.19 ($M_{V2}$) for the secondary component. Adopting the calibrations (April 2022 update) for main sequence stars from \citet{2013ApJS..208....9P} we estimate the mass of the primary as $1.04M_{\odot}$. This value is in good agreement with the \textit{GAIA} estimate of $0.98M_{\odot}$ and $1.0M_{\odot}$ from the TESS input catalogue \citep{2021arXiv210804778P}. Given the agreement between various sources, we adopt the mass $1.01M_{\odot}\pm0.03$ for the primary component. We adopted the combined error of the photometry and distance as the potential error for the mass of the primary component and for the ongoing propagation of errors.

\section{Light Curve Solution and Orbital Stability}
\subsection{Light Curve Solution}
The $V$ band light curve was modelled using a recent version (2015) of the Wilson-Devenney (WD) code \citep{1971ApJ...166..605W, 1990ApJ...356..613W}. The WD code requires the observed curve to be represented as a flux nomalised to the brightest magnitude. The modelled curve is also normalised with maximum brightness at phases 0.25 and 0.75. The standard grid search procedure, suitable in the presence of complete eclipses, was used to find the best fitting solution over a range of mass ratios (0.06 - 15.00). Modelling was performed using mode 3 (overcontact) with the temperature of the secondary ($T_2$), inclination ($i$), potential (fillout) ($f$) and the scaling factor luminosity of the primary ($L_1$) being the adjustable parameters. The limb darkening coefficients were derived from \citet{1993AJ....106.2096V} while the gravity coefficients ($g_1=g_2$) were fixed at 0.32 and the bolometric albedoes ($A_1=A_2$) fixed at 0.5 and simple reflection treatment as applied. The temperature of the primary component is usually fixed during modelling and was set 5600K based on the medium resolution spectrum from the Large Sky Area Multi-Object Fiber Spectroscopic Telescope (LAMOST) survey \citep{2012RAA....12..723Z}. In the case of complete eclipses there is a very strong correlation in the [$q$,$i$] domain such that the absolute value of $T_1$ has little influence on the geometric solution. Variations in $T_1$ of up to 1000K have been modelled yielding the same geometric solution \citep{2023SerAJ.207...21W}. If the temperature of the primary is not used in further analysis, as in the present study, the adopted value will not influence orbital stability analysis. As no significant O'Connell effect (variation in maximum brightness) was observed we did not include star spots in our analysis.

The geometric solution is summarised in Table 1 and the modelled and fitted curves along with a 3D representation of the system are illustrated in Figure 1. Briefly, modelling reveals an extreme low mass ratio system with $q=0.081$ and marginal contact. Our solution is similar to the automated solution of Kepler photometry obtained by \citep{2020PASJ...72..103L}. As noted by \citet{2021PASP..133h4202L}, random errors contribute negligibly to the total error in the case of WD solutions with a low mass ratio completely eclipsing systems, and as such we adopt the standard deviation of the parameters (as reported by the WD code) as the parameter errors. Absolute parameters such as the mass of the secondary component, radii and separation follow from the combination of the geometric solution and Kepler's third law. The absolute parameters are also summarised in Table 1.

\begin{figure}
    \label{fig:fig1}
	\includegraphics[width=\columnwidth]{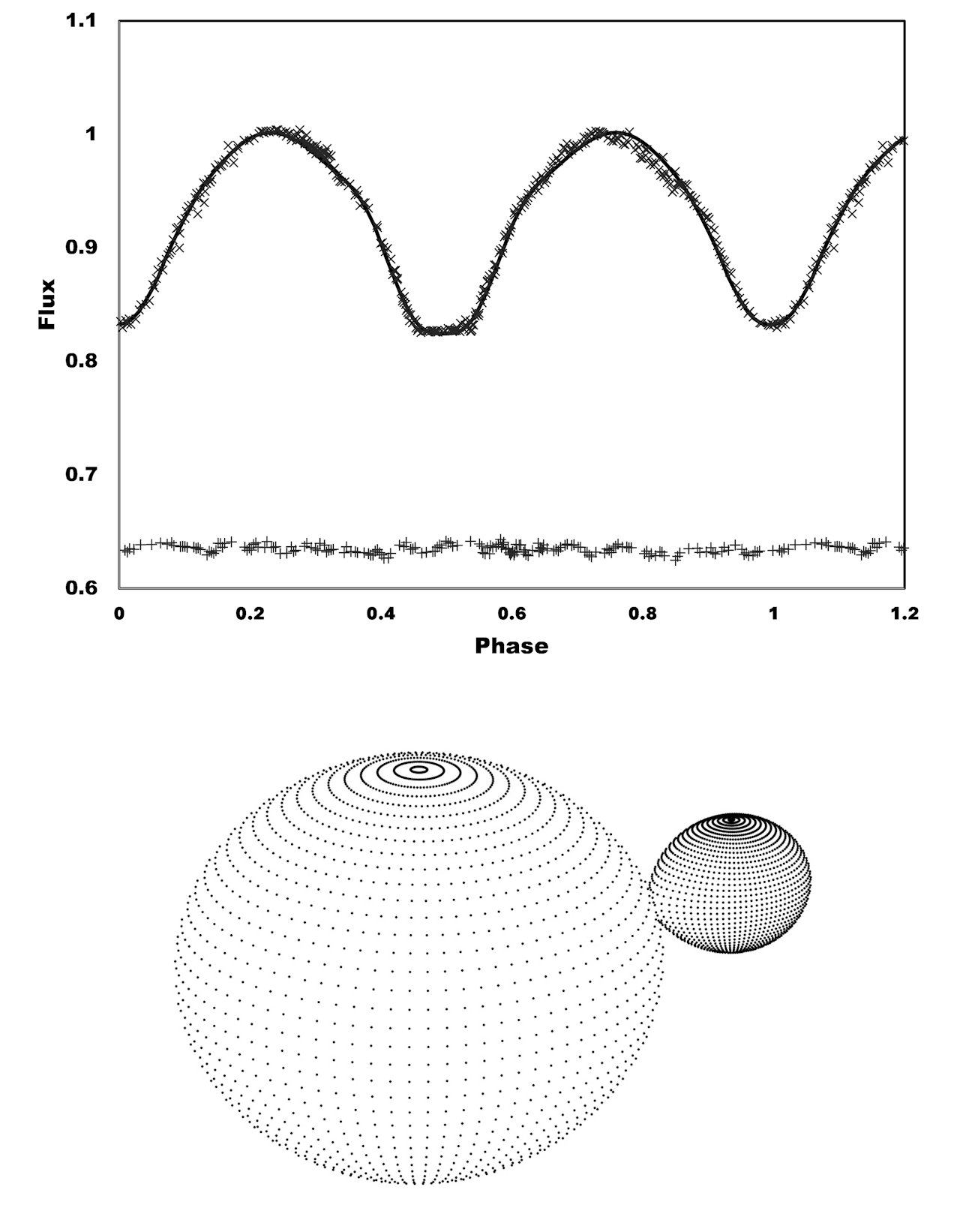}
    \caption{Top: Modelled (Black) and Observed (X) light curves of K9272. Check star flux (+).\\ Bottom: 3D representation of K9272}
    \end{figure}

    \begin{table}

\caption{Light Curve Solution and Absolute Parameters of K9272. SMA = separation\\}
\centering
   \begin{tabular}{|l|l|l|l|}
    \hline
        \hfil $T_1$(K) - Fixed &\hfil 5600 \\
        \hfil $T_2$(K) &\hfil5743$\pm12$\\
        \hfil $q$ &\hfil 0.081$\pm0.001$ \\
        \hfil Incl($^o$) &\hfil 68.5$\pm0.4$\\
        \hfil Fillout (\%) &\hfil 4$\pm2$ \\
        \hfil $M_1/M_{\odot}$ &\hfil  1.01$\pm0.03$\\
        \hfil $M_2/M_{\odot}$&\hfil  0.081$\pm0.004$\\
        \hfil$R_1/R_{\odot}$ &\hfil 1.13$\pm0.02$\\
        \hfil $R_2/R_{\odot}$&\hfil 0.34$\pm0.02$\\
        \hfil SMA ($R_{\odot}$) &\hfil 1.85$\pm0.03$ \\
        \hline
        
    \end{tabular}

    \end{table}

\subsection{Orbital Stability}

Orbital stability of contact binaries has received increasing attention since the confirmation of the red nova V1309 Sco as contact binary merger event. Our research group, over the last few years, has refined the theoretical instability criteria for contact binaries \citep{2021MNRAS.501..229W, 2024SerAJ.208....1A, 2024MNRAS.535.2494W}. Initially we showed that the instability mass ratio is not universal; each system has its own unique instability parameters largely dependent on the mass of the primary component and its gyration radius. Subsequently, we further refined the instability parameters by incorporating the effects of metallicity and age. We essentially concluded that there are likely very few, if any, unstable systems in the solar vicinity. The stability of K9272 was assessed using the instability criteria described in the references cited above.

Our earliest refinements did not incorporate the effects of metallicity or system age with respect to orbital stability \citep{2021MNRAS.501..229W}. Based on these the instability mass ratio range for a contact binary with one solar mass primary would be 0.10 - 0.118. Incorporating the effects of age and metallicity \citet{2024MNRAS.535.2494W}, the instability mass ratio range decreases to 0.09 - 0.101. We adopted the metallicity based on the LAMSOT spectra and GAIA photometry both of which indicate a near solar metallicity for the system. As noted by \citet{2024MNRAS.535.2494W} it is not possible to determine the true age of contact binaries and, as adopted by them, we assumed the primary to be a main sequence star halfway through its main sequence lifespan.

The modelled mass ratio of the system at 0.081 is well below the instability mass ratio range taking full consideration of metallicity and age. Similarly, the separation ($1.82R_{\odot}$) and period ($P=0.28061529d$) are below the theoretical instability onset values.

\section{Discussion and Conclusion}
Potential red nova progenitors are rare. Most reported examples suffer from lingering doubts with respect to the validity of the light curve solution \citep[see][ for some recent examples]{2024MNRAS.535.2494W}. There is a direct correlation between the amplitude of the light curve and the mass ratio of a contact binary system when total eclipses are present \citep{1971ApJ...166..605W, 2022JApA...43...94W}. Unfortunately, the light curve amplitude can be significantly reduced in the presence of a third light leading to a falsely reduced mass ratio being modelled. The effect was dramatically illustrated in the case of CRTS J192848.7‑404555 where the presence of a third light reduced the modelled mass ratio to 0.08 based on survey photometry from the true mass ratio of 0.425 based on higher resolution observations \citep{2023AN....34420066W}.

In the case of K9272 the observed luminosity is in keeping with the estimated spectral mass of the primary component. Many previous examples of extreme low mass ratio contact binaries have luminosity far in excess of that predicted by their estimated mass of the primary component, see \citet{2024MNRAS.535.2494W} for some examples. The GAIA DR3 catalogue does show a companion $\approx$ 2.3 arc seconds from K9272. We could not separate the two in our images however the GAIA estimate of the companion brightness is almost 3 magnitude fainter than the main source and as such its influence on the amplitude of the contact binary light curve is not considered significant.

In addition to a simple line of sight third light, many contact binaries have been shown to be part of multiple star systems. We did not find any improvement in the light curve solution with the addition of a third light however, the potential for a tertiary companion to K9272 has been explored by \citet{2016MNRAS.455.4136B}. They found that there is indeed a likely third companion to K9272. Their analysis suggests that the tertiary companion has a long period ($>1400d$) and that it is unlikely to influence the amplitude of the main contact binary in any meaningful way.

Another issue that has not been addressed in the case of many extreme low mass ratio contact binaries is the mass of the secondary component. In many cases, see \citet{2024MNRAS.535.2494W} for representative examples, the estimated mass of the secondary component falls well below the minimum hydrogen fusion mass of $0.08M_{\odot}$. Based on our results and previous modelling by \citep{2020PASJ...72..103L} the mass of the secondary component is estimated to be above the hydrogen fusion mass and the calculated mass ratio in both solutions is in the instability range.

Current theoretical considerations of orbital stability of low mass contact binaries indicate that most contact binaries, at least in solar vicinity, are likely stable and unlikely to merge and lead to a red nova type transient \citep{2024MNRAS.535.2494W}. The analysis of K9272 presented here is among the most definitive for a contact binary system in an unstable configuration. Given the relative brightness of the system long term high cadence follow up observations are within the reach of modest instruments and encouraged.

\section{Data Availability}

Photometry data are available upon reasonable request to the corresponding author.


\acknowledgements\\
{MG acknowledges funding provided by the Ministry of Science, Technological Development and Innovation of the Republic of Serbia through the contract 451-03-136/2025-03/200002.}\\

{BA acknowledges the funding provided by the Science Fund of the Republic of Serbia through project \#7337 "Modeling Binary Systems That End in Stellar Mergers and Give Rise to Gravitational Waves" (MOBY), by the joint project of the Serbian Academy of Sciences and Arts and Bulgarian Academy of Sciences through the project "Optical search for Galactic and extragalactic supernova remnants" and by the Ministry of Science, Technological Development and Innovation of the Republic of Serbia through the contract \# 451-03-136/2025-03/200104.}\\

\noindent This work has made use of data from the European Space Agency (ESA) mission
{\it Gaia} (\url{https://www.cosmos.esa.int/gaia}), processed by the {\it Gaia}
Data Processing and Analysis Consortium (DPAC,
\url{https://www.cosmos.esa.int/web/gaia/dpac/consortium}). Funding for the DPAC
has been provided by national institutions, in particular the institutions
participating in the {\it Gaia} Multilateral Agreement.\\

\noindent Based on data acquired on the Western Sydney University, Penrith Observatory Telescope. We acknowledge the traditional custodians of the land on which the Observatory stands, the Dharug people, and pay our respects to elders past and present.\\

\noindent This research has made use of the SIMBAD database, operated at CDS, Strasbourg, France.\\

\noindent This work makes use of observations from the Las Cumbres Observatory global telescope network.\\



\newcommand\eprint{in press }

\bibsep=0pt

\bibliographystyle{aa_url_saj}

{\small

\bibliography{sample_saj}
}

\begin{strip}

\end{strip}

\clearpage

{\ }

\clearpage

{\ }

\newpage

\begin{strip}

{\ }






\vskip3mm





















\end{strip}


\end{document}